\documentclass[aip,jcp,reprint]{revtex4-1}
\usepackage[utf8]{inputenc}
\usepackage{color}
\usepackage{graphicx}% Include figure files
\usepackage{dcolumn}% Align table columns on decimal point
\usepackage{bm}% bold math
\usepackage{epsfig}
\usepackage{graphicx}
\usepackage{epstopdf}
\usepackage{amsmath}
\usepackage{amssymb}
\usepackage{braket}
\newcommand{\Tr}{\text{Tr}}
\newcommand{\qnot}{\mathbf{q}_0}
\newcommand{\pnot}{\mathbf{p}_0}
\newcommand{\qnotp}{\mathbf{q}^\prime_0}
\newcommand{\pnotp}{\mathbf{p}^\prime_0}
\newcommand{\pt}{\mathbf{p}_t}
\newcommand{\qt}{\mathbf{q}_t}
\newcommand{\ptp}{\mathbf{p}^\prime_t}
\newcommand{\qtp}{\mathbf{q}^\prime_t}
\newcommand{\Dq}{\mathbf{\Delta}_{q_t}}
\newcommand{\Dp}{\mathbf{\Delta}_{p_t}}
\newcommand{\Dpnot}{\mathbf{\Delta}_{p_0}}
\newcommand{\Dqnot}{\mathbf{\Delta}_{q_0}}
\newcommand{\qin}{\mathbf{q}_i}
\newcommand{\pin}{\mathbf{p}_i}
\newcommand{\mqqf}{\mathbf{M}_{qq}^f}
\newcommand{\mqpf}{\mathbf{M}_{qp}^f}
\newcommand{\mpqf}{\mathbf{M}_{pq}^f}
\newcommand{\mppf}{\mathbf{M}_{pp}^f}
\newcommand{\mqqb}{\mathbf{M}_{qq}^b}
\newcommand{\mqpb}{\mathbf{M}_{qp}^b}
\newcommand{\mpqb}{\mathbf{M}_{pq}^b}
\newcommand{\mppb}{\mathbf{M}_{pp}^b}
\newcommand{\mqq}{\mathbf{M}_{qq}}
\newcommand{\mqp}{\mathbf{M}_{qp}}
\newcommand{\mpq}{\mathbf{M}_{pq}}
\newcommand{\mpp}{\mathbf{M}_{pp}}
\newcommand{\gamz}{\mathbf{\gamma}_0}
\newcommand{\gamt}{\mathbf{\gamma}_t}
\newcommand{\G}{\mathbf{G}}

\newcommand{\pbar}{\mathbf{\bar{p}}}
\newcommand{\qbar}{\mathbf{\bar{q}}}
\newcommand{\dx}{\int\text{d}}
\newcommand{\bc}{\mathbf{c}}
\newcommand{\bz}{\mathbf{z}}
\newcommand{\opA}{\hat{\text{A}}}
\newcommand{\opB}{\hat{\text{B}}}

\begin{document}
\preprint{AIP/123-QED}

\title{Validating and Implementing Modified Filinov Phase 
Filtration in Semiclassical Dynamics}

\author{Matthew S. Church}
\affiliation{Department of Chemistry and Chemical Biology, 
Cornell University, Ithaca, New York, 14853, USA}
\author{Sergey V. Antipov}
\affiliation{Currently at Laboratory of Theoretical Physical Chemistry,
Institut des Sciences et Ing\'{e}nierie Chimiques, \'{E}cole Polytechnique F\'{e}d\'{e}rale 
de Lausanne (EPFL), Lausanne CH-1015, Switzerland}
\author{Nandini Ananth}
\email{na346@cornell.edu}
\affiliation{Department of Chemistry and Chemical Biology, 
Cornell University, Ithaca, New York, 14853, USA}

\date{\today}
\begin{abstract} 
The Mixed Quantum-Classical Initial Value Representation (MQC-IVR) 
is a recently introduced approximate semiclassical (SC) method
for the calculation of real-time quantum correlation 
functions. MQC-IVR employs a modified Filinov filtration 
(MFF) scheme to control the overall phase of the SC 
integrand, extending the applicability of SC methods 
to complex systems while retaining their ability
to accurately describe quantum coherence effects.
Here, we address questions regarding the effectiveness of the MFF
scheme in combination with SC dynamics.
Previous work showed that this filtering scheme is of limited 
utility in the context of semiclassical wavepacket propagation,
but we find the MFF is extraordinarily powerful in the context of 
correlation functions.
By examining trajectory phase and amplitude contributions 
to the real-time SC correlation function in a model system, 
we clearly demonstrate that the MFF serves to reduce noise 
by damping amplitude only in regions of highly oscillatory phase
leading to a reduction in computational effort while retaining 
accuracy.
Further, we introduce a novel and efficient MQC-IVR formulation
that allows for linear scaling in computational cost with 
the total simulation length, a significant improvement over
the more-than quadratic scaling exhibited by the original method.
\end{abstract}

\maketitle
\section{\label{sec:level1}Introduction} 
Semiclassical (SC) approximations to exact, real-time 
quantum correlation functions such as the Double Herman-Kluk 
Initial Value Representation~\cite{hk1} (DHK-IVR)
are unique in their ability to accurately describe 
a wide range of chemical processes~\cite{rev1,rev2,rev3,rev4}
including systems where nuclear and electronic 
quantum coherences play a significant dynamic role~\cite{coh,dhk2} 
such as in the vibrational motion following a Franck-Condon 
excitation,~\cite{franck} or in various nonadiabatic processes.~\cite{ananth,nonad1,nonad2,nonad3,nonad4}
Unfortunately, the computational cost of 
converging the highly oscillatory SC integrand 
scales exponentially with system dimensionality,
limiting the application of these methods to relatively
low-dimensional systems.

Efforts to mitigate the so-called SC `sign' problem 
have led to the development of 
several forward-backward techniques~\cite{fb1,fb2,fb3,fb4,fb5,fb6,fb7} 
that offer a small reduction in computational cost while 
retaining accuracy.
However, the simulation of truly complex systems remains
limited to methods that do not fully describe coherence.
This includes SC methods that employ an additional 
linearization~\cite{lsc1,lsc2}
or related approximations~\cite{related1,related2,fbsd1,fbsd2},
as well as model path-integral based dynamics such 
as Centroid Molecular Dynamics,~\cite{cmd1,cmd2}
and Ring Polymer Molecular Dynamics
and its extensions.~\cite{rpmd1,rpmd2,rpmd3,rpmd4,rpmd5,rpmd6}
More recently, a handful of exact~\cite{qcpi} and 
approximate methods that can account for electronic coherence 
effects~\cite{mvrpmd,jess1,cotton,tao_cc} at short times 
have emerged, however the inclusion of nuclear coherences
in high-dimensional systems remains out of reach.

To address this challenge, two of us recently
introduced a new SC method for the calculation 
of real-time correlation functions: the Mixed Quantum-Classical 
Initial Value Representation (MQC-IVR).~\cite{mqcivr}
This method is derived using the modified 
Filinov filtration (MFF) technique~\cite{filinov1,
filinov2,filinov3,filinov4,filinov5,filinov6,
filinov7,filinov8,filinov9,filinov10} to associate a filtration
parameter with each mode of the system; modulating
these parameters allows us to control the strength of
the filter and hence the extent of quantization of 
an individual mode. As demonstrated in our introductory
paper, for an optimal choice of parameters, MQC-IVR 
correlation functions correctly account for quantum 
coherence effects and require significantly fewer trajectories 
to achieve numerical convergence than standard 
quantum-limit SC methods like the DHK-IVR.

In this paper, we analyze the effectiveness 
of the MFF technique, addressing concerns raised by an 
earlier study that found the MFF to be of limited utility 
in the context of SC wavepacket propagation.~\cite{batista}
Specifically, the study found the MFF to be ineffective for 
both low-dimensional non-chaotic systems, where the amplitude of 
the SC integrand in regions of highly oscillatory phase was
shown to be negligible, and for chaotic systems where the MFF
was necessary to achieve numerical convergence but led to 
loss of valuable phase information.~\cite{batista} 
Here, we undertake a similar analysis of the MQC-IVR correlation
function where phase contributions arise from the net
action of a pair of forward-backward trajectories rather than
the action along a single trajectory. We find that in a simple
low-dimensional system, 
(i) the trajectory pairs with highly oscillatory phase
make numerical convergence difficult while contributing
little to the actual ensemble average,
and 
(ii) the MFF technique successfully 
filters contributions from pairs of trajectories 
that contribute highly oscillatory terms to the MQC-IVR integrand,
significantly reducing computational cost while retaining 
accuracy.
Having validated our filtration scheme, we further 
reformulate the MQC-IVR expression to facilitate a 
novel, highly parallel double-forward (DF) implementation 
that exhibits linear scaling in CPU time with 
simulation length, unlike the more-than quadratic scaling 
exhibited by our original forward-backward (FB) 
implementation.

The paper is organized as follows: 
in Section II, we begin with a brief review of the 
MFF technique and the original MQC-IVR formalism. 
In Section III, we introduce a simple 1D model, provide
implementation details, and present our analysis of the 
phase and amplitude of the MQC-IVR integrand, demonstrating 
the effectiveness of the MFF scheme. 
In Section IV, we introduce our modified
MQC-IVR expression with the novel DF implementation and 
numerically calculate the position correlation function,
showing a two-order of magnitude reduction in CPU time 
for our 1D model system. We conclude in Section V.

%%fakesection: theory
\section{Theory}
\subsection{Modified Filinov Filtration}
The MFF technique~\cite{filinov1,filinov2,filinov3} 
can be used to smooth oscillatory integrals 
of the form,
\begin{align}\label{intI}
I=\int\text{d}\mathbf{z}\;g(\mathbf{z})e^{i\phi\left(\mathbf{z}\right)}
\end{align}
where the functions $g(\mathbf{z})$ and  $\phi(\mathbf{z})$ 
are complex-valued.
Using the MFF to smooth the oscillatory integrand, we obtain
\begin{align}\label{dampF}
 I(\bc)=\dx\bz F(\bz;\bc)g(\bz)e^{i\phi(\bz)},
\end{align}
where the prefactor is defined as 
\begin{align}
F(\bz,\bc)=\det\left|\mathbb{I}+i\bc
\frac{\partial^2\phi}{\partial\bz^2}\right|^\frac{1}{2}
e^{-\frac{1}{2}\frac{\partial\phi}{\partial\bz}^T\bc
\frac{\partial\phi}{\partial\bz}},
\end{align}
and $\mathbb{I}$ is the identity matrix.
In Eq.~(\ref{dampF}), 
the filtering strength is determined by elements 
of the diagonal matrix of tuning parameters, $c_{ii}\geq0$. In the limit
of very small tuning parameters the integrand is unchanged,
\begin{align}
\lim_{\bc\rightarrow 0}I(\bc)=I\label{aa},
\end{align}
and in the limit of very large tuning parameter values 
the MFF is equivalent to a stationary phase approximation,
\begin{align}
\lim_{\bc\rightarrow\infty}I(\bc)=\sum_{j}g(\bz_j)
e^{i\phi(\bz_j)}\left[\det\left|\frac{1}{2\pi i}\frac{\partial^2\phi}{\partial\bz^2}\right|_{\bz=\bz_j}\right]^{-\frac{1}{2}},
\label{bb}
\end{align}
where the subscript $j$ indexes stationary phase points.
Modifying the values of $c_{ii}$ thus controls the 
extent to which the $i^\text{th}$ system mode is filtered, 
and can, ideally, be chosen to facilitate fast numerical 
convergence of the integral without loss of valuable 
phase information.

\subsection{Forward-backward MQC-IVR}
The quantum real-time correlation function between two operators 
$\opA$ and $\opB$ for a system with Hamiltonian $\hat{H}$
is written as 
\begin{align}\label{qcorr}
C_{AB}(t)=\Tr\left[\opA e^{\frac{i}{\hbar}\hat{H}t}\opB e^{-\frac{i}{\hbar}\hat{H}t}  \right].
\end{align}
The corresponding semiclassical DHK-IVR approximation to 
Eq.~(\ref{qcorr}) is a double phase space average 
obtained by replacing both the forward 
and backward time propagators with HK-IVR propagators,~\cite{hk1,filinov10,rev3,rev4}
\begin{align}
\label{dhkivr}
C_{AB}(t)=&\frac{1}{(2\pi\hbar)^{2N}}\dx\pnot \dx\qnot 
\dx\ptp \dx\qtp
\nonumber\\
&\times
\braket{\pnot\qnot|\opA|\pnotp\qnotp}
e^{i[S_t(\pnot,\qnot)+ S_{-t}(\ptp,\qtp)]/\hbar}\nonumber\\
& \times 
\braket{\ptp\qtp|\opB|\pt\qt}
C_t(\pnot,\qnot)
C_{-t}(\ptp,\qtp),
\end{align}
where $N$ is the system dimensionality, 
the forward and backward trajectories have 
initial phase space points, $(\pnot,\qnot)$ and $(\ptp,\qtp)$,
final phase space points, $(\pt,\qt)$ and $(\pnotp,\qnotp)$, 
and classical actions, $S_t$ and $S_{-t}$, respectively.
The position representation of the coherent states 
employed in Eq.~(\ref{dhkivr}) is given by 
\begin{align}\label{wavef}
\braket{\mathbf{x}|\mathbf{p}\mathbf{q}}=
\left(\frac{\det\left|\mathbf{\gamma}\right|}
{\pi^N}\right)^\frac{1}{4}e^{-\frac{1}{2}
(\mathbf{x}-\mathbf{q})^T\mathbf{\gamma}
(\mathbf{x}-\mathbf{q})+
\frac{i}{\hbar}\mathbf{p}^T(\mathbf{x}-\mathbf{q})},
\end{align}
where the diagonal $N\times N$ matrix, $\mathbf{\gamma}$, 
determines the spread of the coherent state in phase space.
Finally, in Eq.~(\ref{dhkivr}), the prefactor for 
the forward trajectory is 
\begin{align}
\nonumber
C_t^2(\pnot,\qnot)=& 
\det\bigg|\frac{1}{2}\bigg[\gamt^\frac{1}{2}
\mqq^f\gamz^{-\frac{1}{2}}+\gamt^{-\frac{1}{2}}
\mpp^f\gamz^\frac{1}{2}\\
&
\left.
-i\hbar\gamt^\frac{1}{2}\mqp^f\gamz^\frac{1}{2}+
\frac{i}{\hbar}\gamt^{-\frac{1}{2}}
\mpq^f\gamz^{-\frac{1}{2}}\right] \bigg|,
\end{align}
where $\mathbf{M}_{\alpha\beta}^f=
\frac{\partial\mathbf{\alpha}_t}
{\partial\mathbf{\beta}_0}$ are the 
forward monodromy matrices, and 
the prefactor for the backward trajectory, $C_{-t}$,
is similarly defined in terms of the 
backward monodromy matrices, $\mathbf{M}_{\alpha\beta}^b=
\frac{\partial\mathbf{\alpha}_0^\prime}
{\partial\mathbf{\beta}_t^\prime}$.

Non-zero phase space displacements between a 
pair of forward and backward trajectories
at time $t$,
\begin{align}\label{disp}
\Dp=\ptp-\pt\;\;\text{and}\;\;\Dq=\qtp-\qt,
\end{align}
leads to a net non-zero action, $S_t+S_{-t}$, 
that contributes to the overall phase of a trajectory pair.
Pairs of trajectories with 
zero phase space displacement have zero net action
and make no contribution 
to the overall phase while widely
diverging trajectory pairs make the semiclassical
integrand in Eq.~(\ref{dhkivr})
highly oscillatory and difficult to numerically 
converge. This motivates the use of a filtering technique
like the MFF to smooth the oscillatory integrand by 
controlling the phase space displacement between 
forward and backward trajectories.

Applying the MFF to the oscillatory 
DHK-IVR integrand in Eq.~(\ref{dhkivr}) yields
the MQC-IVR approximation for a real-time 
correlation function~\cite{mqcivr}
\begin{align}
\label{fbmqcivr}
C_{AB}(t)=& \frac{1}{(2\pi)^{2N}}
\dx\pnot\dx\qnot\dx\Dp\dx\Dq
\nonumber\\
&\times
\braket{\pnot\qnot|\opA|\pnotp\qnotp}
e^{i[S_t(\pnot,\qnot)+S_{-t}(\ptp,\qtp)]}
\nonumber\\
&\times
\braket{\ptp\qtp|\opB|\pt\qt}
D(\pnot,\qnot,\Dp,\Dq,\mathbf{c}_p,\mathbf{c}_q)\nonumber\\
&\times e^{-\frac{1}{2}\Dq^T\mathbf{c}_q\Dq}
e^{-\frac{1}{2}\Dp^T\mathbf{c}_p\Dp},
\end{align}
where we set $\hbar=1$ here and in the 
rest of the manuscript.
In Eq.~(\ref{fbmqcivr}), the prefactor 
$D(\pnot,\qnot,\Dp,\Dq,\mathbf{c}_p,\mathbf{c}_q)$ is a 
determinant of an $N\times N$ matrix containing elements 
of the forward and backward monodromy matrices (full definition provided in Appendix~\ref{Dfb1}), 
and $\mathbf{c}_p$ and $\mathbf{c}_q$ are $N\times N$ diagonal matrices of tuning parameters.
%and the block diagonal $2N\times2N$ matrix of tuning parameters 
%is given by
%\begin{align}
%\label{tunmat}
%\bc=\left(
%\begin{array}{cc}
%	\cq	&	0\\
%	0	& \cp
%\end{array}\right).
%\end{align}
Matrix elements of the tuning parameter $\mathbf{c}_p$ control the extent of momentum 
displacement in a particular mode between trajectory pairs, and 
the elements of $\mathbf{c}_q$ similarly control position displacement.
In the limit $\mathbf{c}_p,\mathbf{c}_q\rightarrow 0$, the MQC-IVR expression
in Eq.~(\ref{fbmqcivr}) becomes identical to the DHK-IVR expression
in Eq.~(\ref{dhkivr}).
As $\mathbf{c}_p,\mathbf{c}_q\rightarrow\infty$, the path displacements 
in Eq.~(\ref{disp}) become zero as does the overall action of the 
forward-backward trajectory pairs. In this limit, the MQC-IVR 
correlation function reduces to the Husimi-IVR,
\begin{align}
\label{husimi}
C_{AB}(t)=&\frac{1}{(2\pi)^N}
\dx\pnot\dx\qnot\nonumber\\
&\braket{\pnot\qnot|\opA
|\pnot\qnot}\braket{\pt\qt|\opB
|\pt\qt},
\end{align}
a classical-limit IVR method that employs 
Husimi functions~\cite{hus1}
of the two operators rather than the Wigner functions 
used in the LSC-IVR formulation.
The Husimi-IVR correlation function involves a classical
average, making it an appealing method for high-dimensional 
systems but fails to account for quantum effects like deep 
tunneling and coherence.
Choosing intermediate values for the matrix elements
of $\mathbf{c}_p$ and $\mathbf{c}_q$ thus allows us to control the extent
of `quantum'-ness incorporated in our simulations,
and choosing different values for different modes allows
us to perform mode-specific quantization within a 
uniform semiclassical dynamic framework.

\section{Analyzing the MFF}
\subsection{Model System}\label{sec:fbmodel}
For clarity, we analyze the effectiveness of the MFF 
using the 1D anharmonic oscillator employed as a model
system in our introductory MQC-IVR paper,~\cite{mqcivr}
\begin{align}\label{1dpot}
V(x)=x^2 -0.1x^3+0.1x^4.
\end{align} 
The initial state is chosen to be 
a coherent state, $\ket{p_iq_i}$, and the corresponding 
wavefunction is given by 
\begin{align}
\Psi(x;t=0)=\left(\frac{\gamma_x}{\pi}\right)^\frac{1}{4}
e^{-\frac{\gamma_x}{2}(x-q_i)^2+ip_i(x-q_i)},
\label{1dinit}
\end{align} 
with $q_i=1$, $p_i=0$, and $\gamma_x=\sqrt{2}$, all 
in atomic units. 
\subsection{FB Implementation} 
We calculate the time-dependent position expectation 
value with $\opA=\ket{p_iq_i}\bra{p_iq_i}$, 
the projection operator corresponding to an initial 
coherent state, and $\opB=\hat{x}$, the position operator.
In the MQC-IVR formulation, this expectation value
is written as~\cite{mqcivr}
\begin{align} 
\label{1davx2} 
\langle x
\rangle_t\;=&\;\frac{1}{4\sqrt{\pi^3}}\int\text{d}p_0\int\text{d}q_0\int\text{d}\Delta_{p_t}\,D_q(p_0,q_0,\Delta_{p_t},c_p)\nonumber\\
&\times
e^{i[S_{t}(p_0,q_0)+S_{-t}(p_t^{'},q_t^{'})]}\braket{p_0q_0|p_iq_i}\braket{p_iq_i|p_0^{'}q_0^{'}}\nonumber\\
&\times q_t e^{-\frac{1}{2}c_p\Delta_{p_t}^2},
\end{align} 
where the prefactor $D_q(p_0,q_0,\Delta_{p_t},c_p)$ 
is provided in Appendix~\ref{Dfb2}. Note that the integral over $\Delta_{q_t}$ has
been evaluated analytically in Eq.~(\ref{1davx2}) 
using the $\gamma_t\rightarrow\infty$ limit.~\cite{mqcivr}

Initial positions, momenta, and the momentum jump are 
sampled from the distribution,
\begin{eqnarray}
\omega(p_0,q_0,\Delta_{p_t};c_p)=F|\braket{p_0q_0|p_iq_i}|^2 
e^{-\frac{1}{2}c_p \Delta_{p_t}^2},
\label{fbsample}
\end{eqnarray} 
with $F=\sqrt{\frac{c_p}{(2\pi)^3}}$ for normalization,
and time-evolved using a fourth-order symplectic 
integrator~\cite{integrator} with a time step 
of $\Delta t\,=\,0.05\,\text{a.u.}$ along 
with the monodromy matrices.
Energy conservation along all trajectories 
is monitored with a tolerance parameter, 
$\epsilon=10^{-4}$, such that 
\begin{align}
\left|1-E(t)/E(0)\right|&<\epsilon.
\end{align}
We also track the Maslov index of the prefactor in order 
to select the correct branch of the complex square root.

\subsection{Phase and Amplitude Analysis} 
\begin{figure}[h!]
\begin{center}
\includegraphics[scale=0.37]{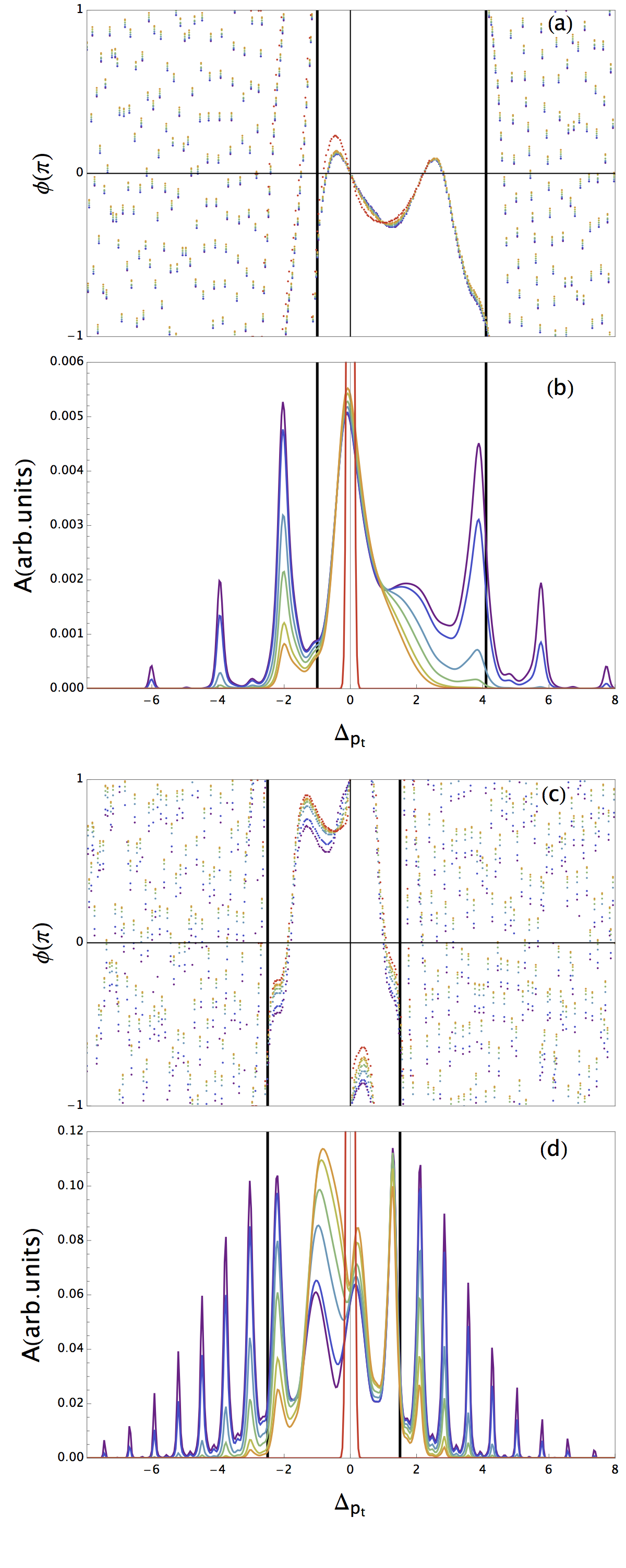}\\
\end{center}
\vspace{-0.5in}
\caption{The contribution to the position correlation 
function from a range of forward-backward trajectory pairs, 
each separated by $\Delta_{p_t}$, and each sharing the same 
initial phase space point for the forward trajectory. 
(a) The integrand phase and (b) the integrand amplitude of 
the MQC-IVR correlation function are plotted against the momentum 
displacement between the forward-backward pair at time 
$t=22\,\text{a.u.}$ 
Figures (c) and (d) show the same quantities
at a later time $t=61\,\text{a.u.}$
The colored lines each correspond to a different Filinov parameter, 
$c_p=0.05$ (purple), $c_p=0.1$ (blue), 
$c_p=0.03$ (blue-green), $c_p=0.5$ (green), 
$c_p=0.7$ (yellow), $c_p=1.0$ (orange), and 
$c_p=200$ (red). 
The vertical black lines in both plots 
enclose the regions of slowly varying 
phase.}
\vspace{-0.2in}
\label{fbint}
\end{figure}

Since the position correlation function defined in 
Eq.~(\ref{1davx2}) involves only a momentum displacement,
$\Delta_{p_t}$, we can analyze the effectiveness of the MFF by 
examining the phase and amplitude contributions to the 
MQC-IVR integrand from different forward-backward trajectory 
pairs that share the same initial conditions at time $t=0$, and 
differ only by the $\Delta_{p_t}$ values.
We choose initial conditions $q_0=1$~a.u. and $p_0=0$~a.u.
corresponding to the center of the initial coherent state, 
propagate forward for time $t$, introduce a change in momentum, 
$\Delta_{p_t}$, and then propagate backward to time $t=0$. 
We plot the amplitude and phase of the integrand 
for each pair of forward-backward trajectories as 
a function of $\Delta_{p_t}$ in Fig.~\ref{fbint}(a) and Fig.~\ref{fbint}(b)
for $t=22$ a.u. and in Fig.~\ref{fbint}(c) and Fig.~\ref{fbint}(d)
for $t=61\,\text{a.u.}$. The two times chosen here correspond
to values at which the classical and quantum 
position correlation functions have significantly 
different amplitudes.

The different colored lines in the phase and 
amplitude plots in Fig.~\ref{fbint} correspond 
to different values of the Filinov parameter, $c_p$, 
ranging from the quantum-limit (shown in purple) 
to the classical-limit (shown in red).
Consider the phase corresponding to the 
momentum jump value $\Delta_{p_t}=0$ 
in Fig.~\ref{fbint}(a) and Fig.~\ref{fbint}(c). 
At this point, the forward and backward trajectories
coincide and the net phase is zero (or $\pm\pi$ due to a phase
contribution from operator $\opB$)  
-- these are the trajectory pairs that 
contribute to the classical-limit 
Husimi-IVR correlation function in Eq.~(\ref{husimi}),
as indicated by the narrow red ($c_p=200$) peak in 
the corresponding amplitude plots in Fig.~\ref{fbint}(b)
and Fig.~\ref{fbint}(d).
As the magnitude of $\Delta_{p_t}$ increases, the path 
of the forward and backward trajectories are significantly
different, resulting in a net non-zero phase contribution 
that varies rapidly with small changes in momentum jump.
Although the regions of slowly varying phase 
in Fig.~\ref{fbint}(a) and Fig.~\ref{fbint}(c) 
(enclosed within solid black lines) are relatively 
invariant to the choice of $c_p$ values, as we go from 
low $c_p$ values that describe 
quantum-limit behavior to high $c_p$ values for classical-limit behavior, 
the amplitude changes significantly, approaching 
zero for large values of $\Delta_{p_t}$. 
Our analysis of the MFF based on a single trajectory 
pair thus suggests that we can eliminate 
contributions from regions of highly oscillatory phase, 
$|\Delta_{p_t}|\ge 3$ in both Fig.~\ref{fbint}(a) and Fig.~\ref{fbint}(c).
%using a filter strength of $c_p\approx 0.7$.

\begin{figure}[h!]
\centering
\includegraphics[scale=0.37]{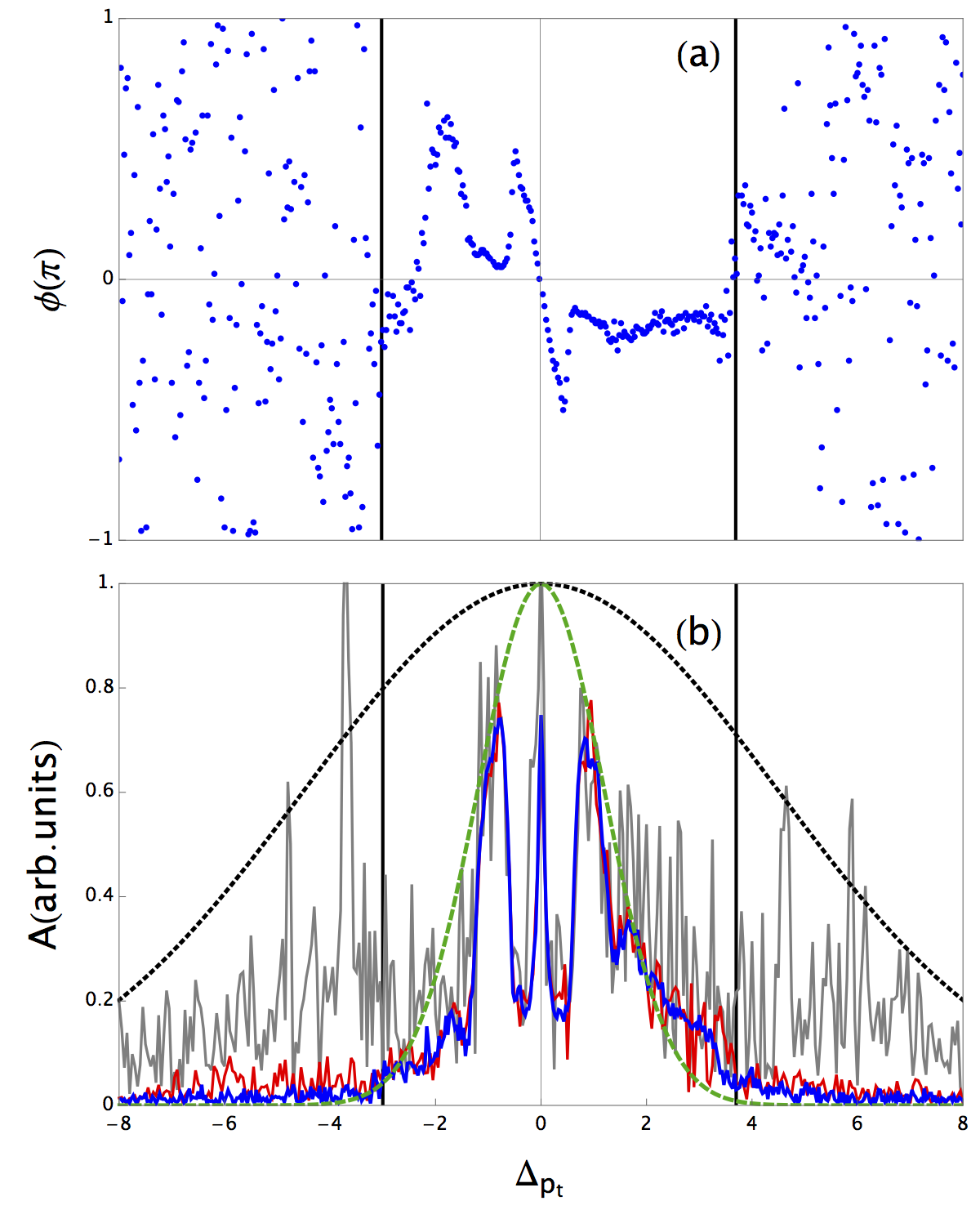}
\vspace{-0.2in}
\caption{The (a) phase and (b) amplitude of the MQC-IVR position correlation function against the
momentum displacement between forward-backward trajectories at $t=22$ a.u. in the
quantum limit, as averaged over $1.2\times10^3$ trajectories (gray), $6.0\times10^4$
trajectories (red), and $2.4\times10^5$ trajectories (blue). The black-dashed Gaussian represents
a weak filter strength $(c_p=0.05)$ while the green-dashed Gaussian represents an optimal
filter strength $(c_p=0.7)$.}
\vspace{-0.1in}
\label{mqcaverage}
\end{figure}

Next we analyze the amplitude and phase of an average integrand 
obtained by summing over contributions from an ensemble of 
forward-backward trajectory pairs to verify that as we approach numerical 
convergence, regions of highly oscillatory phase do, in fact, 
have a negligible contribution to the ensemble average 
correlation function.
Using a weak filter (quantum-limit, $c_p=0.05$) and simulation 
time $t=22$ a.u., we plot the average phase of the MQC-IVR 
integrand in Fig.~\ref{mqcaverage}(a) as a function of the momentum jump.
In Fig.~\ref{mqcaverage}(a) we find a region of slowly varying 
phase corresponding to 
the region $-4$ a.u.$\le\Delta_{p_t}\le 4$ a.u 
(enclosed between two vertical black lines).
The integrand amplitude is plotted in Fig~\ref{mqcaverage}(b);
we find that as the number of trajectories included 
in the ensemble average is increased, the amplitude 
vanishes in regions of highly oscillatory phase. 
Taken together, these observations suggest that trajectory pairs 
with $\Delta_{p_t}$ values outside the slowly varying
phase zone ($|\Delta_{p_t}| \ge 4$ a.u.) contribute very little to the overall
correlation function and make it difficult to achieve numerical
convergence. This makes a strong case for the use of MFF to 
increase computational efficiency -- for instance, with 
a stronger filter, say $c_p=0.7$ (green Gaussian 
in Fig.~\ref{mqcaverage}(b)), it becomes possible to explicitly 
exclude trajectory pairs that contribute only 
noise by making the amplitude zero.
We show, in Fig.~\ref{1Dtcf}(b) of the following section, that an MQC-IVR simulation 
employing $c_p=0.7$ does indeed recover quantum recurrence 
in the position correlation function at long times with 
significantly fewer trajectories.

\section{The Double-Forward MQC-IVR}
\subsection{Re-formulating MQC-IVR}
In the original forward-backward (FB) implementation,~\cite{mqcivr}
with general operators $\opA$ and $\opB$,
we sample the initial phase space points of the forward 
trajectory $(\pnot,\qnot)$ and the path displacement 
variables at time $t$, ($\Dp, \Dq$), to evaluate the integral in 
Eq.~(\ref{fbmqcivr}).
To calculate the MQC-IVR correlation at $N_t$ time points requires
propagating forward trajectories for $N_t$ time steps, 
using the sampled phase space jump values to generate 
new initial conditions for the backward trajectories at each time
step, and then propagating trajectories backward for $N_t$ steps.
Thus, a correlation function with $N_t$ time points involves a total
of $\frac{1}{2}(N_t^2+3N_t)$ propagation steps for each trajectory pair.

Here, we introduce a novel double-forward (DF) implementation 
that significantly reduces the computational effort involved 
in an MQC-IVR calculation.
We re-write the MQC-IVR correlation function 
in Eq.~(\ref{fbmqcivr}) such that the variables of integration 
are the initial phase space points of two independent forward
trajectories,
\begin{eqnarray}
\nonumber
C_{AB}(t)&=&\frac{1}{(2\pi)^{2N}}
\int\text{d}\pnot\int\text{d}\qnot
\int\text{d}\pnotp\int\text{d}\qnotp\\
\nonumber
&\times &\braket{\mathbf{p}_0\mathbf{q}_0|
\hat{\text{A}}|\mathbf{p}_0^\prime\mathbf{q}_0^\prime}
\braket{\mathbf{p}_t^\prime\mathbf{q}_t^\prime
|\hat{\text{B}}|\mathbf{p}_t\mathbf{q}_t}\\
\nonumber
&\times& e^{i[S_t(\pnot,\qnot)-S_t(\pnotp,\qnotp)]}\,
D_t(\pnot,\qnot,\pnotp,\qnotp;\mathbf{c}_p,\mathbf{c}_q)\\ 
&\times&
e^{-\frac{1}{2}\Dqnot^T\mathbf{c}_q\Dqnot}e^{-\frac{1}{2}\Dpnot^T\mathbf{c}_p\Dpnot}.
\label{ffmqcivr}
\end{eqnarray} 
Details of the derivation are provided 
in Appendix~\ref{ff_derive} along with the 
explicit form of the prefactor, $D_t$.
As expected, the limits of the MQC-IVR correlation 
function remain the same as before yielding the DHK-IVR 
when $\mathbf{c}_p,\mathbf{c}_q\rightarrow0$ and the Husimi-IVR when
$\mathbf{c}_p,\mathbf{c}_q\rightarrow\infty$.
We evaluate the correlation function defined 
in Eq.~(\ref{ffmqcivr}) by sampling 
the variables $(\pnot,\qnot)$ and $(\pnotp,\qnotp)$ 
from a correlated sampling distribution,\cite{tao}
\begin{eqnarray} 
\omega(\pnot,\qnot,&\pnotp&,\qnotp;\mathbf{c}_p,\mathbf{c}_q)\label{DFsample}\\
\nonumber
&=& F|\braket{\pbar\,\qbar|\pin\qin}|^2
e^{-\frac{1}{2}\Dpnot^T\mathbf{c}_p\Dpnot}
e^{-\frac{1}{2}\Dqnot^T\mathbf{c}_q\Dqnot},
\end{eqnarray} 
where we define mean variables,
$\mathbf{\bar{x}}=\frac{\mathbf{x}_0'+\mathbf{x}_0}{2}$, 
and we introduce trajectory displacement variables 
at time $t=0$,
$\Dpnot=\pnotp-\pnot$ and $\Dqnot=\qnotp-\qnot$.
For normalization we have $F=\frac{\sqrt{\det|\mathbf{c}_p|\det|\mathbf{c}_q|}}{(4\pi^2)^N}$.
Calculating the MQC-IVR correlation function now 
involves propagating two independent trajectories 
forward in time. The contribution to a correlation
function of length $N_t$ is obtained simply from a 
given pair of forward trajectories using a 
total of $2N_t$ propagation steps.

Thus, the DF implementation exhibits
linear scaling in CPU time with total simulation length
relative to the more-than quadratic scaling of the 
original FB formulation. 
Furthermore, the use of two independently propagated
forward trajectories, rather than a forward-backward 
structure, allows for highly parallel implementation.

%\subsection{Model Systems}
%Model I is the 1D anharmonic oscillator described previously
%in Section~\ref{sec:fbmodel}. 
%Model II is a previously studied 2D system~\cite{mqcivr} 
%where the anharmonic oscillator in Eq.~(\ref{1dpot}) is linearly 
%coupled to a heavy harmonic mode,
%\begin{align}
%V(x,y)=x^2-0.1x^3+0.1x^4+\frac{1}{2}k_yy^2 + fxy,
%\label{2dpot}
%\end{align} 
%where $k_y=25/9$ a.u. and coupling constant $f=2$ a.u. 
%The initial coherent state of this system is 
%the product of two coherent states,
%$\ket{p_xq_x}\ket{p_yq_y}$.
%The corresponding initial wavefunction given is by
%\begin{align} 
%\nonumber
%\Psi(x,y;t=0)=\left(\frac{\gamma_x\gamma_y}{\pi^2}
%\right)^\frac{1}{4}&
%e^{-\frac{\gamma_x}{2}(x-q_x)^2+ip_x(x-q_x)}\\
%&\times e^{-\frac{\gamma_y}{2}(y-q_y)^2+ip_y(y-q_y)},
%\end{align} 
%where the subscript $x/y$ denotes $x/y$-component of a vector,
%the anharmonic $x$-mode parameters are as specified 
%for the 1D model previously, and the harmonic mode 
%parameters in atomic units are $q_y=1$, $p_y=0$, and 
%$\gamma_y=25/3$.

\subsection{Simulation Details}
Initial conditions for the forward 
trajectory pairs are sampled using the correlated sampling function 
in Eq.~(\ref{DFsample}) with $N=1$.
As before, we calculate the position correlation function, 
$\langle x\rangle_t$, where $\opA=\ket{p_iq_i}\bra{p_iq_i}$
and $\opB=\hat{x}$
using the DF implementation in Eq.~(\ref{ffmqcivr}).
The matrix element of the position operator is given by 
\begin{equation}\label{poselem1D}
    \braket{\ptp \qtp|\hat{x}|\pt\qt}=\frac{1}{2}\big[(q_{tx}^\prime+q_{tx})-\frac{i}{\gamma_{xx}}\Delta_{p_{tx}}\big]\braket{\ptp\qtp|\pt\qt}.
\end{equation} 
Details of trajectory integration remain the same as before. 
The SC-Corr code package,~\cite{corrcode}
developed in-house and available as open-source 
software, was used to perform these calculations.
%\nacomment{Please read carefully and make sure we are not 
%missing any information that would be necessary for someone else
%to implement}

\subsection{Numerical Results for DF Correlation Functions}

\begin{figure}[h!]
\begin{center}
\includegraphics[scale=0.27]{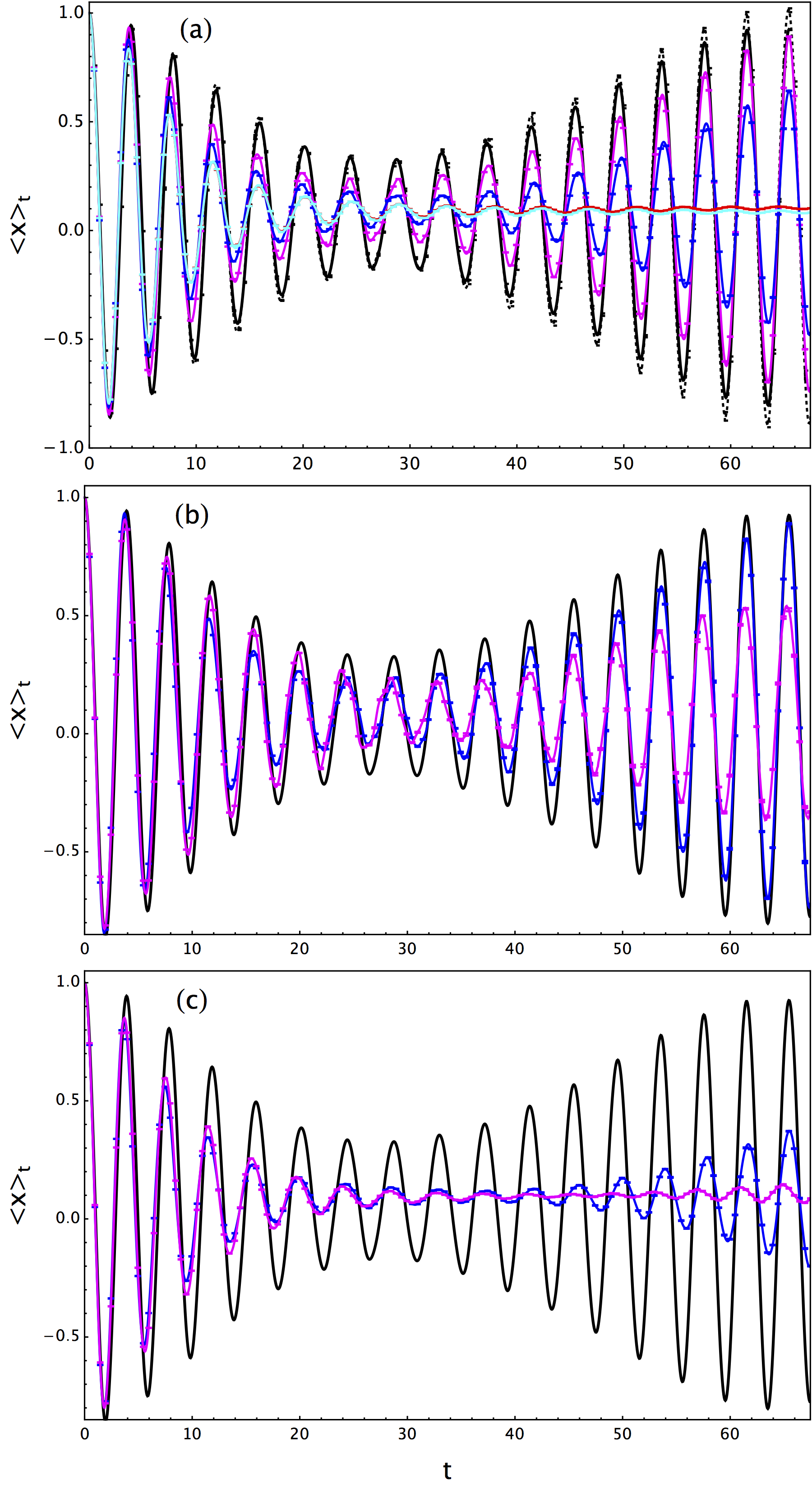}
\end{center}
\vspace{-0.3in}
\caption{The position correlation function of the 1D system with 
(a) exact quantum (black), DHK-IVR (dotted), Husimi-IVR (cyan), 
and the DF implementation of MQC-IVR 
with $c=0.7$ (magenta), $c=3.0$ (blue), $c=500$ (red).
The lower two figures compare results 
obtained from the original FB implementation
of MQC-IVR (magenta) and the new DF implementation of
MQC-IVR (blue) with (b) $c=0.7$ and (c) $c=10$. 
The exact quantum result is shown in black.}
%\vspace{-0.2in}
\label{1Dtcf}
\end{figure}

In Fig.~\ref{1Dtcf}(a), we plot 
$\langle x\rangle_t$ as a function of time obtained from the DF
implementation with different values of the tuning 
parameters $c_q=c_p=c$. Comparing this result
against exact quantum, DHK-IVR, and Husimi-IVR 
simulations,
we demonstrate that the new DF implementation, like the 
original FB implementation,~\cite{mqcivr} 
recovers the amplitude recurrences at long times 
that distinguish the quantum result from the 
classical result. 
As the filter strength ($c$ value) 
is increased, we obtain classical-limit results
where the amplitude is damped at long times.
In Table~\ref{1dtab}, we report 
the number of trajectories required 
to converge each simulation such that the maximum 
absolute statistical error across all 
time steps is $\le 5\%$. 
Even for a quantum-limit filter strength, 
$c=0.7$, the DF implementation captures quantum 
behavior with two orders of magnitude fewer 
trajectories than required by a full semiclassical 
DHK-IVR calculation.
We note that the number of trajectories required 
to achieve numerical convergence with the DF
implementation is comparable to the original FB
implementation, but the CPU time required 
per trajectory pair is significantly lower.

\begin{table}[h!] 
\centering 
\begin{tabular}{ c c c c c c c}
\hline 
IVR formulation & $\;\;\;$ & $\;\;\;$ & $c$ & $\;\;\;$ & $\;\;\;$ &
$N_{traj}$ \\\hline\hline
DHK & $\;\;\;$ & $\;\;\;$ & 0 & $\;\;\;$ & $\;\;\;$
& $3.0\times10^6$ \\ \hline & $\;\;\;$ & $\;\;\;$ & $0.7$ & $\;\;\;$ & $\;\;\;$
& $2.4\times10^4$ \\ DF-MQC & $\;\;\;$ & $\;\;\;$ & $3.0$ & $\;\;\;$ & $\;\;\;$ &
$9.6\times10^3$ \\ & $\;\;\;$ & $\;\;\;$ & $500.0$ & $\;\;\;$ & $\;\;\;$ &
$6.0\times10^2$ \\ 
\hline Husimi & $\;\;\;$ & $\;\;\;$ & $\infty$ & $\;\;\;$ &
$\;\;\;$ & $2.4\times10^2$ \\ 
\hline 
\end{tabular} 
\caption{The total number of trajectories, $N_{traj}$, required to achieve
numerical convergence for the position correlation function, 
as computed with different semiclassical formulations.}
\label{1dtab} 
\vspace{-0.2in}
\end{table} 
An additional advantage of the DF implementation
becomes apparent in Fig.~\ref{1Dtcf}(b) and Fig.~\ref{1Dtcf}(c)
where we compare the position correlation function obtained 
from the DF and FB MQC-IVR implementations
for $c=0.7$ and $c=10$ respectively. 
The DF formulation better captures quantum 
recurrences at long times since the filter
now determines phase space displacements at time 
zero, allowing a broader range of trajectory
displacements to be included at time $t$.
Further, the total CPU time required 
to compute $\langle x \rangle_t$ with $c=0.7$, 
$2.4\times10^4$ trajectories, and $1600$ time steps 
for both implementations is reported in Table~\ref{time},
demonstrating that the DF implementation increases 
computational efficiency by two orders of magnitude.

\begin{table}[h!] 
\centering 
\begin{tabular}{ c c c c c c c }
\hline%\hline
Implementation & $\;\;\;$ & $\;\;\;$ & $\;\;\;$ & CPU Time /seconds & $\;\;\;$ &
$\;\;\;$ \\ 
\hline DF & $\;\;\;$ & $\;\;\;$ & $\;\;\;$ & $2$ &
$\;\;\;$ & $\;\;\;$ \\ 
FB & $\;\;\;$ & $\;\;\;$ & $\;\;\;$ &
$288$ & $\;\;\;$ & $\;\;\;$ \\ 
\hline%\hline 
\end{tabular} 
\caption{
	A comparison of CPU time to calculate the MQC-IVR position
correlation function with the two implementations and $c=0.7$.
%\nacomment{I changed this to minutes - please make sure 
%I converted correctly!}
}
\label{time} 
\end{table} 

\section{Conclusions} 
We show that the MFF technique 
used in MQC-IVR acts to {\it apriori}
filter out phase contributions from widely
diverging trajectory pairs that contribute
little, on average, to the correlation function. 
This makes it an effective and efficient tool for enhancing 
numerical convergence in the calculation of 
correlation functions without 
loss of important quantum information.

Further, we introduce a novel DF implementation 
for the MQC-IVR correlation function that scales 
linearly with simulation length compared to 
the quadratic scaling of the FB implementation.
This, combined with an improved ability to capture
quantum recurrences at long times, immediately
extends the applicability of MQC-IVR to 
high-dimensional system simulations.

\section*{Acknowledgements}
This work is supported in part by the Army Research 
Office Grant No. W911NFD-13-1-0102 and in part by Cornell University 
Start-Up funding. N. A. additionally acknowledges 
support from the Research Corporation for Science through a Cottrell 
Scholar Award and a Sloan Foundation Research Fellowship.

%%fakesection: appendix
\appendix
\section{Forward-backward MQC-IVR prefactors}
\subsection{General operator $\opB$} \label{Dfb1} 
The MQC-IVR prefactor for a general
operator $\opB$ is given by,
\begin{eqnarray}
&& D(\pnot,\qnot,\Dp,\Dq;\mathbf{c}_p,\mathbf{c}_q)=
2^{-\frac{N}{2}}\det(\gamz^{-1}\G)^\frac{1}{2}\nonumber\\
&&\times\det\bigg[\frac{1}{2}(\mppb-i\gamz\mqpb)(\G^{-1}+
    \mathbb{I})(\mppf\gamz+i\mpqf)\nonumber\\
&&+(\gamz\mqqb+i\mpqb)(\frac{1}{2}\gamt^{-1}+\mathbf{c}_p)
\G^{-1}(\mppf\gamz+i\mpqf)\nonumber\\
&&+\frac{1}{2}(\gamz\mqqb+i\mpqb)(\G^{-1}+\mathbb{I})
(\mqqf-i\mqpf\gamz)\nonumber\\
&&+(\mppb-i\gamz\mqpb)(\frac{1}{2}\gamt+\mathbf{c}_q)
\G^{-1}(\mqqf-i\mqpf\gamz)  \bigg]^{\frac{1}{2}},\nonumber
\end{eqnarray}
with diagonal matrix $\G=(\mathbf{c}_q+\gamt)\mathbf{c}_p+\mathbf{c}_q(\gamt^{-1}+\mathbf{c}_p)$. The superscript $f$ denotes
the monodromy matrix element corresponding to the forward trajectory beginning
at $(\pnot,\qnot)$, while the superscript $b$ denotes the monodromy matrix element
corresponding to the backward trajectory beginning at $(\ptp,\qtp)$.

\subsection{Position-space operator $\opB=\opB(\hat{q})$}
The MQC-IVR prefactor for a position-space operator $\opB(\hat q)$ is 
obtained by evaluating the $\gamma_t\rightarrow\infty$ limit,
\label{Dfb2}
\begin{eqnarray}
&&D_q(\pnot,\qnot,\mathbf{\Delta}_{p_t},\mathbf{c}_p)\;=\;\nonumber\\
&&\det\bigg|\gamz^{-1}\mathbf{c}_p\bigg[(\mppb-i\gamz\mqpb)(\mppf\gamz+i\mpqf)\nonumber\\
&&+\;(\gamz\mqqb+i\mpqb)(\mqqf-i\mqpf\gamz)\nonumber\\
&&+\;(\mppb-i\gamz\mqpb)\mathbf{c}_p^{-1}(\mqqf-i\mqpf\gamz)\bigg]\bigg|^\frac{1}{2}\nonumber
\end{eqnarray}
\section{Double Forward MQC-IVR Formulation}
\subsection{Changing variables of integration}\label{ff_derive}
We start with the DHK-IVR of Eq.~(\ref{dhkivr}) 
and change variables of integration from $(\ptp,\qtp)$ 
to $(\pnotp,\qnotp)$ using Liouville's theorem. 
The action of the backward trajectory, $S_{-t}(\ptp,\qtp)$, is then 
replaced by its forward counterpart, $S_{t}(\pnotp,\qnotp)$, and 
we use the following monodromy matrix identity for the backward
trajectory in order to obtain a 
new prefactor expression,
\begin{eqnarray}
\mathbf{M}^{b}\;=\;(\mathbf{M}^{f\prime})^{-1}\;=\;
\begin{pmatrix}
\mathbf{M}_{pp}^{f^T\prime}	&	-\mathbf{M}_{qp}^{f^T\prime}\\
-\mathbf{M}_{pq}^{f^T\prime}	&	\mathbf{M}_{qq}^{f^T\prime}
\end{pmatrix}\nonumber,
\end{eqnarray}
with $\mathbf{M}_{\alpha\beta}^{f\prime}=\frac{\partial \alpha_t^\prime}{\partial\beta_0^\prime}$
%where \nacomment{define the Mqq etc for the new forward 
%    version of the backward explicitly otherwise it may be 
%    confused with the regular forward trajectory Monodromy matrix
%elements in the text.}
The resulting expression for the correlation function is a 
double-forward DHK-IVR,
\begin{align}
C_{AB}(t)=&\frac{1}{(2\pi)^{2N}}
\dx\pnot\dx\qnot\dx\pnotp\dx\qnotp\nonumber\\
&\times C_t(\pnot,\qnot)C_{t}(\pnotp,\qnotp)
e^{i[S_t(\pnot,\qnot)-S_{t}(\pnotp,\qnotp)]}\nonumber\\
&\times\braket{\pnot\qnot|\hat{\text{A}}|\pnotp\qnotp}
\braket{\ptp\qtp|\hat{\text{B}}|\pt\qt}.
\label{ffdhkivr}
\end{align}
We now follow the original MQC-IVR derivation,~\cite{mqcivr} 
implementing the MFF scheme to smooth the oscillatory
integrand in Eq.~(\ref{ffdhkivr}). 
We take $\phi$ to include the coherent state exponentials
in Eq.~(\ref{ffdhkivr}) and the action terms, excluding any contributions 
from operators $\opA$ and $\opB$ as well as 
any phase contributions from the prefactors,
\begin{eqnarray}
\phi&\;=\;&S_t(\pnot,\qnot)\;-\;S_t(\pnotp,\qnotp)+\frac{i}{4}\Delta_{q_0}^T\gamz\Delta_{q_0}+\frac{i}{4}\Delta_{p_0}^T\gamz^{-1}\Delta_{p_0}\nonumber\\
&&-\frac{1}{2}(\pnotp+\pnot)^T\Delta_{q_0}+\frac{i}{4}\Delta_{q_t}^T\gamt\Delta_{q_t}+\frac{i}{4}\Delta_{p_t}^T\gamt^{-1}\Delta_{p_t}\nonumber\\
&&+\frac{1}{2}(\ptp+\pt)^T\Delta_{q_t}\nonumber.
\end{eqnarray}
The derivation proceeds much like the original method 
(from Eq.~(12) onward in the original manuscript~\cite{mqcivr}), 
but the vector of phase space displacements is now given by
\begin{eqnarray}
\mathbf{y}\;=\;
\begin{pmatrix}
\Delta_{q_0}\\
\Delta_{p_0}\\
\Delta_{q_t}\\
\Delta_{p_t}
\end{pmatrix}\nonumber,
\end{eqnarray}
and the block diagonal matrix of tuning parameters is
\begin{eqnarray}
\bc\;=\;
\begin{pmatrix}
\;\mathbf{c}_q\;	&	\;\;0\;\;	&	\;\;0	&	\;\;0\\
\;0\;\,	&	\;\mathbf{c}_p\;	&	\;\;0	&	\;\;0\\
\;0\;\,	&	\;\;0\;\;	&	\;\;0	&	\;\;0\\
\;0\;\,	&	\;\;0\;\;	&	\;\;0	&	\;\;0
\end{pmatrix}\nonumber.
\end{eqnarray}
After simplifying the integrand we obtain Eq.~(\ref{ffmqcivr}) and 
the prefactor takes the form,
\begin{eqnarray}
&&D(\pnot,\qnot,\pnotp,\qnotp;\mathbf{c}_p,\mathbf{c}_q)\;=\;\det(\frac{1}{2}\gamt^{-1}\G)^\frac{1}{2}\nonumber\\
&&\times\det\bigg[\frac{1}{2}(\mppf-i\gamt\mqpf)(\G^{-1}+\mathbb{I})(\mppb\gamt+i\mpqb)\nonumber\\
&&+(\gamt\mqqf+i\mpqf)(\frac{1}{2}\gamz^{-1}+\mathbf{c}_p)\G^{-1}(\mppb\gamt+i\mpqb)\nonumber\\
&&+\frac{1}{2}(\gamt\mqqf+i\mpqf)(\G^{-1}+\mathbb{I})(\mqqb-i\mqpb\gamt)\nonumber\\
&&+(\mppf-i\gamt\mqpf)(\frac{1}{2}\gamz+\mathbf{c}_q)\G^{-1}(\mqqb-i\mqpb\gamt)\bigg]^\frac{1}{2}\nonumber,
\end{eqnarray}
with diagonal matrix $\G=(\mathbf{c}_q+\gamz)\mathbf{c}_p+\mathbf{c}_q(\gamz^{-1}+\mathbf{c}_p)$.
We note that unlike the FB implementation, 
the DF implementation
will typically always include both position 
and momentum jumps
even if operator $\opB$ is a pure 
position or momentum operator.

%%fakesection: bibliography 

%%% fakesection: Things to fix throughout manuscript
%\newpage
% \noindent
%\nacomment{
%    1.Citation after punctuation throughout manuscript.\\
%    2.All equation reference should be standardized -- put equation
%    number in parantheses or not depending on JCP format.\\
%    3. The figures are all slightly off-center. This is likely because
%    there is extra white space to the left -- you want to screenshot such
%    that the whitespace on both sides is the same and also minimal.\\
%    4. Read carefully for: grammar, all quantities in all equations must be fully defined, where equations are references make sure they are the right ones, and carefully re-check equations.
%    5. Make sure figure captions are all complete
%    6. When done with this list, delete this section so there is nothing
%    after the bibliography.
%}

\begin{thebibliography}{500} 
%% review articles forintroduction 
%Herman-Kluk
\bibitem{hk1}M. F. Herman and E. Kluk, Chem. Phys. \textbf{91}, 27 (1984).
\bibitem{rev1}W. H. Miller, J. Phys. Chem. A \textbf{105}, 2942 (2001). 
\bibitem{rev2}M. Thoss and H. Wang, Annu. Rev. Phys. Chem. \textbf{55}, 299 (2004).
\bibitem{rev3}K. G. Kay, J. Chem. Phys. \textbf{100}, 4377 (1994).
\bibitem{rev4}K. G. Kay, Ann. Rev. Phys. Chem. \textbf{56}, 255 (2005).
\bibitem{coh} W. H. Miller, J. Chem. Phys. \textbf{136}, 210901 (2012).
\bibitem{dhk2}J. M. Moix and E. Pollak, Phys. Rev. A. \textbf{79}, 062507 (2009).
\bibitem{franck} H. Wang, M. Thoss, K. Sorge, R. Gelabert, X. Gim\'{e}nez, and
W. H. Miller, J. Chem. Phys. \textbf{114}, 2562-2571 (2001).
\bibitem{ananth} N. Ananth, C. Venkataraman, and W. H. Miller, J. Chem. Phys. \textbf{127}, 084114 (2007).
\bibitem{nonad1} X. Sun and W. H. Miller, J. Chem. Phys. \textbf{106}, 6346 (1997).
\bibitem{nonad2} X. Sun, H. Wang and W. H. Miller, J. Chem. Phys. \textbf{109}, 7064 (1998).
\bibitem{nonad3} M. Thoss, G. Stock and W. H. Miller, J. Chem. Phys. \textbf{112}, 10282 (2000).
\bibitem{nonad4} E. A. Coronado, V. S. Batista and W. H. Miller, J. Chem. Phys. \textbf{112}, 5566 (2000).
\bibitem{fb1} N. Makri and K. Thompson, Chem. Phys. Lett. \textbf{291}, 101 (1998).
\bibitem{fb2} X. Sun and W. H. Miller, J. Chem. Phys. \textbf{110}, 6635 (1999).
\bibitem{fb3} W. H. Miller, Faraday Discuss. Chem. Soc. \textbf{110}, 1 (1998).
\bibitem{fb4} H. Wang, M. Thoss, and W. H. Miller, J. Chem. Phys. \textbf{112}, 47 (2000).
\bibitem{fb5} J. Shao and N. Makri, J. Phys. Chem. A \textbf{103}, 7753 (1999).
\bibitem{fb6} R. Gelavert, X. Gim\'{e}nez, M. Thoss, H. Wang and W. H. Miller, J. Chem. Phys. \textbf{114}, 2572 (2001).
\bibitem{fb7} K. Thompson and N. Makri, Phys. Rev. E. \textbf{59}, R4729, (1999).
\bibitem{lsc1} H. Wang, X. Sun, and W. H. Miller, J. Chem. Phys. \textbf{108}, 9726 (1998). 
\bibitem{lsc2} X. Sun, H. Wang, and W. H. Miller, J. Chem. Phys. \textbf{109}, 4190 (1998). 
\bibitem{related1} J. Shao and N. Makri, J. Phys. Chem. A \textbf{103}, 9479 (1999).
\bibitem{related2}S. J. Cotton and W. H. Miller, J. Phys. Chem. A \textbf{117}, 7190 (2013).
\bibitem{fbsd1}N. Makri, J. Phys. Chem. B \textbf{106}, 8390 (2002).
\bibitem{fbsd2}J. Kegerreis and N. Makri, J. Comp. Chem. \textbf{28}, 818 (2007).
\bibitem{cmd1}J. Cao and G. A. Voth, J. Chem. Phys. \textbf{100}, 5106 (1994). 
\bibitem{cmd2}S. Jang and G. A. Voth, J. Chem. Phys. \textbf{111}, 2371 (1999).
\bibitem{rpmd1}I. R. Craig and D. E. Manolopoulos, J. Chem. Phys. \textbf{121}, 3368 (2004). 
\bibitem{rpmd2}S. Habershon, D. E. Manolopoulos, T. E. Markland, and T. F. Miller III, 
Annu. Rev. Phys. Chem. \textbf{64}, 387 (2013). 
\bibitem{rpmd3}A. R. Menzeleev, N. Ananth, and T. F. Miller III, J. Chem. Phys. \textbf{135}, 074106 (2011). 
\bibitem{rpmd4} T. J. H. Hele, M. J. Willatt, A. Muolo and S. C. Althorpe, J. Chem. Phys. \textbf{142}, 134103 (2015).
\bibitem{rpmd5}A. R. Menzeleev, F. Bell, and T. F. Miller III, J. Chem.
\bibitem{rpmd6}J. R. Duke and N. Ananth, Faraday Discuss. \textbf{195}, 253 (2016).
\bibitem{qcpi} N. Makri, Int. J. Quantum Chem. \textbf{115}, 1209 (2015).
\bibitem{mvrpmd} N. Ananth, J. Chem. Phys. \textbf{139}, 124102 (2013).
\bibitem{jess1} J. R. Duke and N. Ananth, J. Phys. Chem. Lett. \textbf{6}, 4219 (2015).
\bibitem{cotton} S. J. Cotton and W. H. Miller, J. Chem. Phys. \textbf{139}, 234112 (2013).
\bibitem{tao_cc} G. Tao, J. Phys. Chem. Lett. \textbf{7}, 4335 (2016).
\bibitem{mqcivr}S. V. Antipov, Z. Ye, and N. Ananth, J. Chem. Phys. \textbf{142}, 184102 (2015).
\bibitem{filinov1}V. S. Filinov, Nucl. Phys. B \textbf{271}, 717 (1986).
\bibitem{filinov2}N. Makri and W. H. Miller, Chem. Phys. Lett. \textbf{139}, 10 (1987).
\bibitem{filinov3}N. Makri and W. H. Miller, J. Chem. Phys. \textbf{89}, 2170 (1988).
\bibitem{filinov4}B. W. Spath and W. H. Miller, J. Chem. Phys. \textbf{104}, 95 (1996).
\bibitem{filinov5}B. W. Spath and W. H. Miller, Chem. Phys. Lett. \textbf{262}, 486 (1996). 
\bibitem{filinov6}M. F. Herman, Chem. Phys. Lett. \textbf{275}, 445 (1997). 
\bibitem{filinov7}E.A. Coronado, V.S. Batista, and W.H. Miller, J. Chem. Phys. \textbf{112}, 5566 (2000).
\bibitem{filinov8}A. R. Walton and D. E. Manolopoulos, Mol. Phys. \textbf{87}, 961 (1996). 
\bibitem{filinov9}X. Sun and W. H. Miller, J. Chem. Phys. \textbf{108}, 8870 (1998). 
\bibitem{filinov10} M. Thoss, H. Wang, and W. H. Miller, J. Chem. Phys. \textbf{114}, 9220 (2001). 
%\bibitem{dhk1}S. Zhang and E. Pollak, J. Chem. Phys. \textbf{121}, 3384 (2004).
\bibitem{batista}M. Spanner, V. S. Batista, and P. Brumer, J. Chem. Phys. \textbf{122}, 084111 (2005).
\bibitem{hus1} K. Husimi, Proc. Phys. Math. Soc. Jpn. {\bf 122}, 264  (1940).
\bibitem{integrator}M. L. Brewer, J. S. Hulme, and D. E. Manolopoulos, J. Chem. Phys. \textbf{106}, 4832 (1997). 
\bibitem{tao} F. Pan and G. Tao, J. Chem. Phys. \textbf{138}, 091101 (2013).
\bibitem{corrcode}https://github.com/AnanthGroup/SC-IVR-Code-Package
\end{thebibliography}
\end{document}